# Spin transfer in an antiferromagnet

Z. Wei[1], A. Sharma[2], A. S. Nunez[1], P. M. Haney[1], R. A. Duine[1], J. Bass[2], A. H. MacDonald[1], and M. Tsoi[1]

[1]*Physics Department, University of Texas at Austin, Austin, Texas 78712, USA*
[2]*Department of Physics and Astronomy, Michigan State University, East Lansing, Michigan 48824, USA*

**An electrical current can transfer spin angular momentum to a ferromagnet[1-3]. This novel physical phenomenon, called spin transfer, offers unprecedented spatial and temporal control over the magnetic state of a ferromagnet and has tremendous potential in a broad range of technologies, including magnetic memory and recording. Recently, it has been predicted[4] that spin transfer is not limited to ferromagnets, but can also occur in antiferromagnetic materials and even be stronger under some conditions. In this paper we demonstrate transfer of spin angular momentum across an interface between ferromagnetic and antiferromagnetic metals. The spin transfer is mediated by an electrical current of high density (~$10^{12}$ A/m$^2$) and revealed by variation in the exchange bias at the ferromagnet/antiferromagnet interface. We find that, depending on the polarity of the electrical current flowing across the interface, the strength of the exchange bias can either increase or decrease. This finding is explained by the theoretical prediction that a spin polarized current generates a torque on magnetic moments in the antiferromagnet. Current-mediated variation of exchange bias can be used to control the magnetic state of spin-valve devices, e.g., in magnetic memory applications.**

Spin valves[5] are now used in magnetic field sensors, in read heads for hard drives, in galvanic isolators, and in non-volatile random access memory devices. The simplest type of spin valve consists of two ferromagnetic layers separated by a thin nonmagnetic spacer. The spin-valve resistance is smallest when the magnetizations of the two ferromagnetic layers are parallel and largest when the magnetizations are antiparallel. The antiparallel alignment is achieved by making the two layers respond differently to an external magnetic field; an antiferromagnet in contact with one of the layers is used to effectively 'pin' the magnetization in this layer through an effect called 'exchange bias'[6-8]. The exceptional responsiveness of spin valves to magnetic fields has enabled very high areal packing densities in hard drives. In our experiments we study how exchange bias behaves when extremely high current densities are driven across these spin valve structures using point contacts.

Point contacts were instrumental both for the original observation of spin transfer in ferromagnetic materials[3] and in probing the high-frequency manifestation of this phenomenon[9-11]. The extremely small size, less than a trillionth of a square cm, qualifies point contact as the smallest probe of spin transfer phenomena today and enables current densities up to $10^{13}$ A/m$^2$. Our point contacts were made with a standard system[3,12], using a sharpened Cu wire and a differential screw mechanism to move the Cu tip toward a FeMn/CoFe/Cu/CoFe spin valve structure. The spin-valve structures were sputtered onto Si substrates with individual layer thicknesses from 3-10 nm using techniques already described[13], capped with a Au protective layer, and had a thick Cu underlayer. The latter was used to secure a perpendicular-to-plane flow of the current (CPP) from the point contact, across the spin valve, and into the Cu buffer. All measurements reported here were made on a FeMn (8 nm)/CoFe (3 nm)/Cu (10 nm)/CoFe (10 nm) spin valves with a 5 nm thick Au protective layer and a 50 nm thick Cu buffer. The samples were cooled through the Néel temperature of FeMn ($T_N \approx 400$ K) in the presence of a static magnetic field (~18mT) to pin the magnetization direction of the neighboring CoFe. Inverted structures where the pinning antiferromagnet (FeMn) lies on top of the spin-valve stack were also tested.

At room temperature and in magnetic fields B up to 0.1 T applied along the exchange bias direction, we have measured the magnetoresistance (MR) of point contacts at different bias currents. Figure 1 shows typical variations in the contact resistance R=V/I as a function of the applied field B (solid traces) for a series of bias currents I. Black (grey) traces show sweeps from high positive (negative) field to high negative (positive) fields. Here positive current flows from the tip into the spin valve. For a given bias current I (given trace in Fig. 1) the form of R(B) is typical for spin valves: starting from high positive field, R(B) is constant at a minimum value (magnetizations of the two CoFe layers are parallel), rises to a maximum when the magnetization of top (free) CoFe layer switches near zero field (leading to antiparallel alignment of the two CoFe layers), and then decreases to its minimum value beyond the exchange bias field at which the magnetization of the pinned CoFe is finally reversed. The reversed sweeps from high negative to high positive fields show similar behavior. The reversal of both free and pinned CoFe layers and the corresponding variations in R proceed via a discrete series of steps. The latter correspond to reversals of individual ferromagnetic domains in CoFe probed by the point contact. Note that domains closest to the contact contribute the most to its resistance.

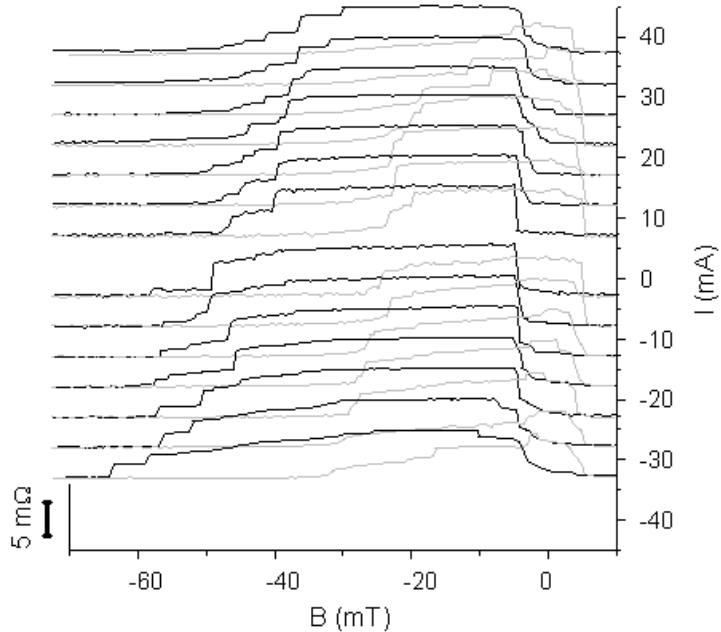

**Figure 1 Point-contact magnetoresistance at different bias currents.** Solid traces show point-contact resistance R=V/I as a function of the applied magnetic field B for a series of bias currents

I (-35, -30, -25, -20, -15, -10, -5, 5, 10, 15, 20, 25, 30, 35 mA). Black (grey) traces are for B-sweeps down (up). Note that the resistance changes associated with reversal of the pinned layer mostly extend over a wider range of B (broader transition) than those for the free layer, especially for larger (magnitude) I. Point-contact resistance at high fields is 0.92 Ω.

The reversal of the free layer seems to be little affected by the applied current. In contrast, the current clearly changes the exchange bias field at which the pinned layer is reversed. The 2D gray-scale plot representation (see Fig. 2a; lighter color indicates higher resistance) of the data in Fig. 1 suggests that on average the exchange bias increases (decreases) with applied negative (positive) current. The white lines in Fig.2 are guides to the eye which emphasize the overall trend, and also emphasize stochastic variations that occur on top of this trend. Although the resistance curve near switching varies from run-to-run at a given current, the typical behavior point-contact GMR behavior, the trend indicated by the white lines is always present. For comparison, Fig. 2b shows exchange bias variations in an inverted structure where the pinning antiferromagnet lies on top of the spin-valve stack CoFe (10nm)/Cu (10nm)/ CoFe (3nm)/FeMn (8nm). Here positive current crosses the FeMn/CoFe interface in the opposite sense (from FeMn into CoFe) compared to the case of Fig. 2a. Interestingly, the effect of current on the exchange bias is also reversed – the positive (negative) current increases (decreases) the exchange bias.

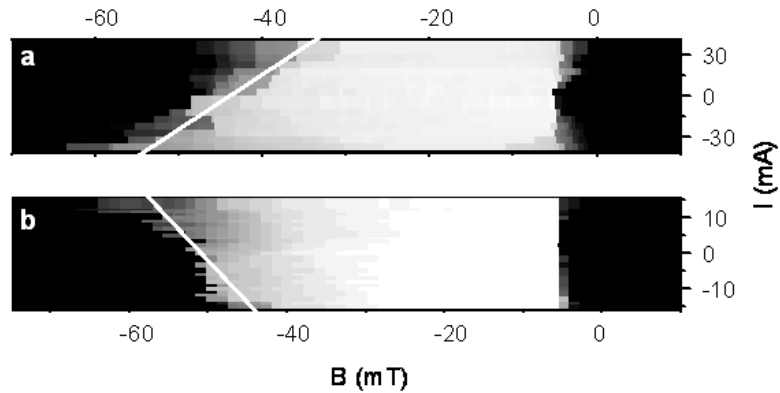

**Figure 2 Variation of exchange bias in standard and inverted spin-valve structures. a,** 2D gray-scale plot representation of the down sweeps from Fig. 1 (lighter color indicates higher resistance) for a standard spin-valve structure (FeMn8nm/CoFe3nm/Cu10nm/CoFe10nm). The positive (negative) current *decreases* (increases) the exchange bias (see white line for guide). **b,** 2D gray-scale plot of the current-dependent magnetoresistance in an inverted spin-valve structure where the pinning antiferromagnet lies on top of the spin-valve stack (CoFe10nm/Cu10nm/CoFe3nm/FeMn8nm). Here positive (negative) current *increases* (decreases) the exchange bias. The point-contact resistance at high fields is 1.59 Ω.

To interpret our data we use the ideas put forward in refs. 4 and 14 in which torques are calculated by evaluating electron spin densities in the non-equilibrium current-carrying state. In this time-dependent mean-field (e.g. time-dependent spin-density-functional) based theory spin-torques are due to the contribution of transport electrons near the Fermi energy to the spin-dependent exchange-correlation potential, which is[6] in turn proportional to the corresponding spin-density contribution. Current- induced changes in the exchange-correlation effective magnetic field are experienced by all quasiparticles. Orbitals well away from the Fermi energy, which behave collectively, experience the changes as a spin-torque that is in general non-zero in both ferromagnets and antiferromagnets. Conservation of total spin, which is relevant for order parameter dynamics in a ferromagnet but not in an antiferromagnet, is not a necessary condition for current-induced spin-torques but can reduce its description to the commonly used action-reaction picture involving transport and collective orbitals.

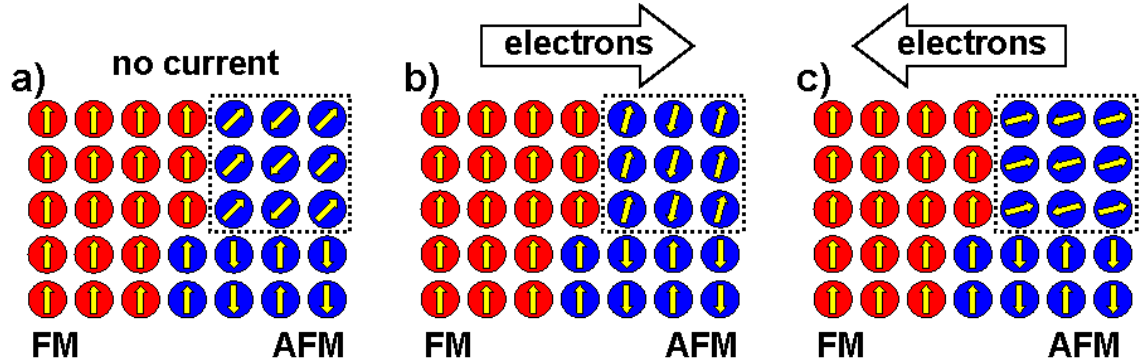

**Figure 3 Schematic illustration of the influence of transport currents on exchange bias.** In the absence of a current, the moment-orientation variation in the surface layer of the ferromagnet is highly non-uniform but has a non-zero average value established by annealing in the presence of a magnetic field. In the presence of a transport current, spin-torques in the antiferromagnet alter its moment orientation distribution and can either increase (b) or decrease (c) the surface layer polarization and the exchange bias.

Our explanation for the dependence of exchange bias on transport currents that we have discovered is summarized schematically in Fig. 3. Near the exchange bias field, the meta-stability of the ferromagnet's *opposite to field* orientation is due almost entirely to exchange interactions with the surface layer of the antiferromagnet. The exchange bias field is always much smaller than the ferro/antiferro exchange interaction because of surface and magnetic roughness at the interface. When electrons flow into the antiferromagnet from the ferromagnet, its moments experience torques that drive the surface layer orientation toward that of the ferromagnet. As illustrated in panel (b) above, this change in the antiferromagnet's moment orientation configuration will increase the polarization of its surface layer (for ferromagnetic interactions across the interface) and therefore increase the exchange bias. A current flowing in the opposite direction [panel (c)] will have the opposite effect on the surface layer polarization and hence on the exchange bias. The sense of the antiferromagnetic spin-torque follows from scattering state solutions[14] of the Schroedinger equation for generic toy-model ferromagnet/antiferromagnet interfaces, and should be thought of as a rule of thumb rather than a theorem. Importantly, this torque acts throughout the antiferromagnet[4] and not just on its surface layer.

Although this simple picture does not attempt to account in detail for the (likely complicated) domain structure in both materials near the FeMn/CoFe interface that is presumably responsible for stochastic run-to-run variations, we believe that it gives the correct qualitative explanation for the overarching trend we have indentified in our data. Because there is no spacer separating the ferromagnet and the antiferromagnet, strong interactions across the interface imply that changes in the magnetic microstructure of the antiferromagnet will not normally be permanent, but will relax once the current is turned off. Our results suggest though, that it is possible in principle to drive irreversible changes in the antiferromagnet's microstructure with a transport current, and thereby achieve post-growth changes in exchange-bias characteristics.

**References:**

1. Slonczewski, J. C. Current-driven excitation of magnetic multilayers. *J. Magn. Magn. Mater.* **159**, L1-L7 (1996).
2. Berger, L. Multilayer as spin-wave emitting diodes. *J. Appl. Phys.* **81**, 4880-4882 (1997).
3. Tsoi, M. *et al.* Excitation of a magnetic multilayer by an electric current. *Phys. Rev. Lett.* **80**, 4281-4284 (1998).
4. Núñez, A. S., Duine, R. A., Haney, P. M., and MacDonald, A. H. Theory of spin torques and giant magnetoresistance in antiferromagnetic metals. *Preprint cond-mat*/0510797, to appear in *Phys. Rev. B*.
5. Dieny, B. *et al.* Giant magnetoresistive in soft ferromagnetic multilayers. *Phys. Rev. B* **43**, 1297–1300 (1991).
6. Meiklejohn, W. H. & Bean, C. P. New magnetic anisotropy. *Phys. Rev.* **102**, 1413-1414 (1956).
7. Nogués, J. & Schuller, I. K. Exchange bias. *J. Magn. Magn. Mater.* **192**, 203-232 (1999).
8. Berkowitz, A. E. & Takano, K. Exchange anisotropy - a review. *J. Magn. Magn. Mater.* **200**, 552-570 (1999).
9. Tsoi, M. *et al.* Generation and detection of phase-coherent current-driven magnons in magnetic multilayers. *Nature* **406**, 46-48 (2000).
10. Krivorotov, I. N. *et al.* Time-domain measurements of nanomagnet dynamics driven by spin-transfer torques. *Science* **307**, 228-231 (2005).
11. Kaka, S., *et al.* Mutual phase-locking of microwave spin torque nano-oscillators. *Nature* **437**, 389-392 (2005).
12. Jansen, A. G. M., van Gelder, A. P. & Wyder, P. Point-contact spectroscopy in metals. *J. Phys. C* **13**, 6073-6118 (1980).
13. Slaughter, J. M., Pratt, W. P., Jr & Schroeder, P. A. Fabrication of layered metallic systems for perpendicular resistance measurements. *Rev. Sci. Instrum.* 60, 127-131 (1989).
14. Núñez, A. S. & MacDonald, A. H. Spin transfer without spin conservation. *Preprint cond-mat*/0403710, to appear in *Solid State Commun*.